# Comparing Measured Fluorocarbon Leader Breaking Strength with Manufacturer Claims

Christine Haight[1], Kadie McNamara[1], Kathleen McQueeney[1], and Ya'el Courtney[2]

[1] U.S. Air Force Academy, 2354 Fairchild Drive, USAF Academy, CO 80840

[2] BTG Research, P.O. Box 62541, Colorado Springs, CO, 80962

## Abstract

The experiment reported in this article addresses manufacturer claims of fluorocarbon leader material strength versus experimental tests of leaders strength. Breaking strength of fishing line is the most common specification when marketing fishing line. In this study, eight leaders rated near 15 pounds by their manufacturers were tested. Each leader was tested with a knot in the line and without a knot in the line. The strongest leader tested without a knot was Cabela's Seaguar fluorocarbon and the weakest leader tested without a knot was Cabela's Premier leader. The highest strength of leaders with a knot was the Ande Monofilament Fluorocarbon and the lowest breaking strength of leaders with a knot was the Seaguar Grand Max Fluorocarbon. Few published studies actually test the breaking strength of a leader to determine the accuracy of manufacturers' claims. Tensile strengths are also reported.

## Introduction

The purpose of this project was to determine if manufacturers' claims of fishing leader strength were accurate. Eight different brands and types of leaders were tested. Each leader was made of fluorocarbon, a polymer that is supposed to be denser, more abrasion resistant and thinner than monofilament leaders. Fluorocarbon is also known for stretching very little and handling greater weight. Many anglers prefer fluorocarbon because its index of refraction is much closer to that of water than ordinary monofilament, making it nearly invisible to fish. The eight types of leaders tested were Yo Zuri Fluorocarbon (15lb), Frog Hair FC Fluorocarbon (12lb), Cabela's Seaguar Fluorocarbon (15lb), Cabela's Premier Fluorocarbon (15lb), Ande Monofilament Fluorocarbon (15lb), Pline CFX Fluorocarbon (15lb), Maxima Fluorocarbon (15lb), and Seaguar Grand Max Fluorocarbon (14.5lb). It was hypothesized that some fishing leaders would be below their manufacturer claims and few would be above. Each leader was tested with and without a knot in the line. It was hypothesized that the knot would decrease the strength of the fishing leader.

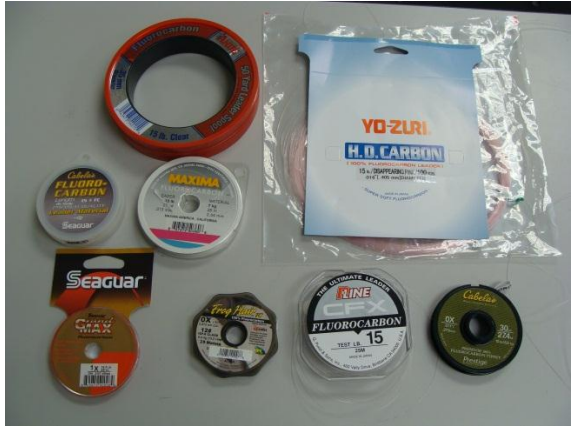

*Figure 1: The eight brands and types of line tested in the experiment.*

**Method**

Fluorocarbon leader materials were tested using a force plate with a circular wooden handle on each end of the line. The leader was wrapped around the lower part of the handle connected to the force plate and then wrapped around the top part of the second circular handle which then hangs below. Transparent tape was then used to secure the leader to the handle. After placing the line for the test, the computer is then activated with the data collection software program, Logger Pro, which is used to calculate the force in newton's versus time in seconds. The control of the experiment was that the same individual exerted the force on the lowest handle, pulling the line until its breaking point and that same individual tied the knot. Each line type was tested with five trials with a Uni Knot and five trials without a knot. Results from each trial were entered into Microsoft Excel and the force was converted from newtons to pounds. Results from all five trials were then averaged to compute the average breaking strength of each line type. The uncertainty, percent of rating, and tensile strength for each leader was also computed.

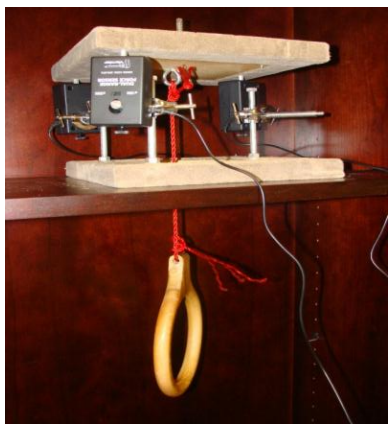

*Figure 2: Force plate used for measuring force in newtons of breaking strength of fishing line. Another handle is added below the handle pictured and is pulled to exert force on the line.*

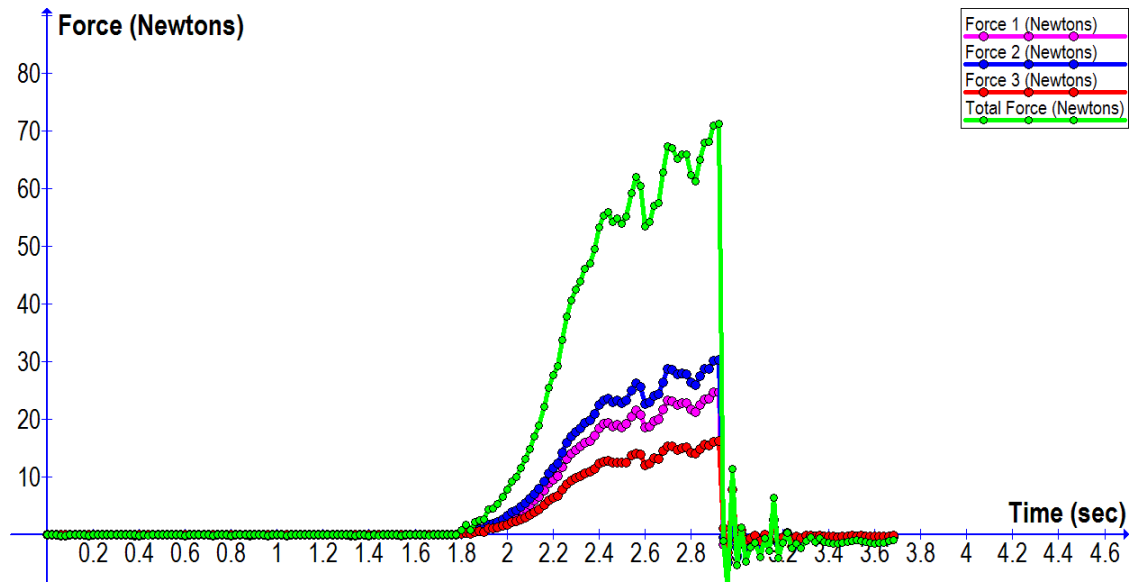

*Figure 3: The resulting graph of force vs. time using Logger Pro software. The graph above is the Maxima Fluorocarbon 15lb line which had a breaking strength of 71.29 newtons which is equal to 16.32 pounds, 1.32 pounds above the manufacturer's claim. Each of the bottom three traces is the force on one of the three force transducers in the force plate. The top trace is the sum, which is the total force on the plate exerted by the fishing line.*

| Fishing Line Brand/Type | Rated Strength (lbs) | Averag Breaking Strength (lbs) | Uncertainty (lbs) | Percent of Rating % | Diameter (in) | Tensile Strength (psi) |
|---|---|---|---|---|---|---|
| Cabelas Seguar Fluorocarbon | 15 | 19.48 | 0.25 | 130 | 0.013 | 146700 |
| Ande Monofilament Fluorocarbon | 15 | 17.62 | 0.16 | 117 | 0.013 | 132700 |
| Frog Hair FC Fluorocarbon | 12 | 13.77 | 0.19 | 115 | 0.011 | 144900 |
| Maxima Fluorocarbon | 15 | 16.46 | 0.25 | 110 | 0.013 | 124000 |
| Pline CFX | 15 | 15.24 | 0.44 | 102 | 0.013 | 114800 |
| Yo Zuri | 15 | 14.6 | 0.3 | 97 | 0.016 | 72600 |
| Seguar Grand Max Fluorocarbon | 14.5 | 13.88 | 0.1 | 96 | 0.01 | 176700 |
| Cabelas Premier | 15 | 12.36 | 0.27 | 82 | 0.011 | 130100 |

*Table 1: Average breaking strength, uncertainty, and percent of rating of unknotted leader materials.*

**Results**

Table 1 shows the average breaking strength of each leader material without a knot. Table 2 shows the average breaking strength of each leader material with a knot. The average breaking strength of each line tested with a knot was lower than the line without the knot, thus supporting the hypothesis that a knot would weaken the line. The knot affected the Seaguar Grand Max Fluorocarbon line the most averaging a breaking strength of 9.98 pounds with a knot, 3.99 pounds less than the average breaking strength without a knot. The Frog Hair FC Fluorocarbon, Ande Monofilament Fluorocarbon and Maxima Fluorocarbon, each with a knot, still had an average breaking strength above the manufacturers claim for that line.

| Fishing Line | Rated Strength | Average Breaking Strength | Uncertainty | Percent of Rating | Diameter | Tensile Strength |
|---|---|---|---|---|---|---|
| Brand/Type | (lbs) | (lbs) | (lbs) | % | (in) | (psi) |
| Ande Monofilament Fluorocarbon | 15 | 17.46 | 0.34 | 116 | 0.013 | 131600 |
| Frog Hair FC Fluorocarbon | 12 | 13.49 | 0.41 | 112 | 0.011 | 142000 |
| Cabelas Seguar Fluorocarbon | 15 | 16.07 | 2.05 | 107 | 0.013 | 121100 |
| Maxima Fluorocarbon | 15 | 15.52 | 0.19 | 103 | 0.013 | 116900 |
| Pline CFX | 15 | 13.07 | 0.88 | 87 | 0.013 | 98470 |
| Yo Zuri | 15 | 12.5 | 0.11 | 83 | 0.016 | 62100 |
| Cabelas Premier | 15 | 10.55 | 0.86 | 70 | 0.011 | 111100 |
| Seguar Grand Max Fluorocarbon | 14.5 | 9.98 | 0.25 | 69 | 0.01 | 127000 |

*Table 2: Knotted breaking strength, uncertainty and percent of rating. These lines were all tested with a Uni Knot. The diameter of Cabelas Seaguar leader and Pline CFX leader were not advertised by the manufacturer and were measured using calipers.*

With the results of the experiment, Cabela's Seaguar Fluorocarbon proves to be the strongest fishing line, exceeding manufacturers' claims by 4.48 pounds and a percent of rating of 129.85%. Ande Monofilament Fluorocarbon and Frog Hair FC Fluorocarbon are also higher ranking for line strength as both have a percent of rating over 110%. Although Cabela's Seaguar is tested as the strongest leader, Cabela's Premier 15 pound fishing line tests weakest with the biggest difference in manufacturers claim and tested result. Cabela's Premier average breaking strength rated at 12.36 pounds, 2.64 pounds under its rated strength. Yo Zuri and Seaguar Grand Max Fluorocarbon also resulted with the average breaking strength under the rated strength ranging from a .4 pound to .7 pound difference. Although the Ande Monofilament Fluorocarbon does not test with the highest average breaking strength, its breaking strength is above the manufacturer's claim with a Uni Knot, (2.46 pounds over rated strength), and without a Uni Knot, (2.62 pounds over rated strength.) The percent of rating was just 1.03% different without a knot tied. Therefore, although a knot in the leader weakens the breaking strength, there was little change between the Ande Monofilament Fluorocarbon without a knot and the Ande Monofilament Fluorocarbon with a knot.

**Conclusion**

The force plate combined with Logger Pro software was successful in measuring the breaking strength of fishing line, and a spreadsheet was useful in computing the uncertainty, and percent of rating. This experiment supported the hypothesis that a knot in fishing line decreases the breaking strength for all lines tested, as most lines with a knot had a significantly lower breaking point than the line without a knot. The experiment also supported the hypothesis that manufacturers' claims vary in their accuracy. Since Uni Knots are used with leaders while fishing, the overall recommended leader was the Ande Monofilament Fluorocarbon, because it exceeded the manufacturer's claims by 2.46 pounds with a Uni Knot. The weakest leader was the Seaguar Grand Max Fluorocarbon which tested 4.52 pounds below the manufacturer's claims with a Uni Knot. With any study, there is the possibility for weaknesses. Possible errors in this study could be contributed to knots not being tied uniformly. Also, our sample size was small in that we only tested one spool of line of each brand and type. Possibilities exist for future work with this project in which multiple spools of each leader type could be tested and averaged together instead of depending on one spool to be the average or normal quality the company produces. There is a large amount of leader types and brands used for fishing and more brands with different rated strengths could also be tested in the future.

**Acknowledgements**
This research was funded by BTG Research (www.btgresearch.org). Advice and technical assistance was provided by Dr. Michael Courtney.